\documentclass[12pt]{article}
\usepackage{amsmath,amsfonts,amssymb,makeidx,ifthen}
\usepackage[dvipdfmx]{graphicx}
\makeatletter
\long\def\@makecaption#1#2{{\small
\advance\leftskip1cm
\advance\rightskip1cm
\vskip\abovecaptionskip
\sbox\@tempboxa{#1: #2}%
\ifdim \wd\@tempboxa >\hsize
 #1: #2\par
\else
\global \@minipagefalse
\hb@xt@\hsize{\hfil\box\@tempboxa\hfil}%
\fi
\vskip\belowcaptionskip}}
\makeatother
\def\eq#1\en{\begin{equation}#1\end{equation}}  
\def\eqa#1\ena{\begin{align}#1\end{align}}
\def\eqg#1\eng{\begin{gather}#1\end{gather}}
\newcommand{\lb}[1]{\label{e:#1}}
\newcommand{\rlb}[1]{\eqref{e:#1}} 
\newcommand{\nl}{\notag\\}

\newcommand{\bkt}[1]{\left\langle#1\right\rangle}
\newcommand{\sbkt}[1]{\langle#1\rangle}
\newcommand{\bbkt}[1]{\bigl\langle#1\bigr\rangle}

\newcommand{\sumtwo}[2]%
{\mathop{\sum_{#1}}_{#2}}
\newcommand{\sumthree}[3]%
{\mathop{\mathop{\sum_{#1}}_{#2}}_{#3}}
\newcommand{\sumfour}[4]%
{\mathop{\mathop{\mathop{\sum_{#1}}_{#2}}_{#3}}_{#4}} 
\newcommand{\prodtwo}[2]%
{\mathop{\prod_{#1}}_{#2}}
\newcommand{\mintwo}[2]%
{\mathop{\min_{#1}}_{#2}}
\newcommand{\maxtwo}[2]%
{\mathop{\max_{#1}}_{#2}}
\newcommand{\maxthree}[3]%
{\mathop{\mathop{\max_{#1}}_{#2}}_{#3}}
\newcommand{\limtwo}[2]%
{\mathop{\lim_{#1}}_{#2}}
\newcommand{\suptwo}[2]%
{\mathop{\sup_{#1}}_{#2}}
\newcommand{\supthree}[3]%
{\mathop{\mathop{\sup_{#1}}_{#2}}_{#3}}
\newcommand{\supfour}[4]%
{\mathop{\mathop{\mathop{\sup_{#1}}_{#2}}_{#3}}_{#4}} 
\newcommand{\inftwo}[2]%
{\mathop{\inf_{#1}}_{#2}}
\newcommand{\infthree}[3]%
{\mathop{\mathop{\inf_{#1}}_{#2}}_{#3}}
\newcommand{\inffour}[4]%
{\mathop{\mathop{\mathop{\inf_{#1}}_{#2}}_{#3}}_{#4}} 

\newcommand{\mrx}{\mathrm{x}}
\newcommand{\mry}{\mathrm{y}}
\newcommand{\mrz}{\mathrm{z}}

\newcommand{\bbC}{\mathbb{C}}

\newcommand{\bbR}{\mathbb{R}}

\newcommand{\ep}{\varepsilon}
\newcommand{\up}{\uparrow}
\newcommand{\dn}{\downarrow}

\newcommand{\Di}{\mathit{\Delta}}
\newcommand{\qedm}{\rule{1.5mm}{3mm}}

\newcommand{\bigno}{\par\bigskip\noindent}

\newcommand{\ket}[1]{|#1\rangle}
\newcommand{\DE}{\Di E}

\newcommand{\hc}{\hat{c}}
\newcommand{\hcd}{\hat{c}^\dagger}
\newcommand{\hn}{\hat{n}}
\newcommand{\hT}{\hat{T}}
\newcommand{\hTd}{\hat{T}^\dagger}
\newcommand{\hV}{\hat{V}}
\newcommand{\rl}{\hat{\rho}_\ell}
\newcommand{\Drl}{\Di\hat{\rho}_\ell}

\newcommand{\bkGS}[1]{\langle\Phi_L^{\rm GS}|#1|\Phi_L^{\rm GS}\rangle}

\newcommand{\HL}{\hat{H}_L}
\newcommand{\Wl}{\hat{W}_\ell}
\newcommand{\Ul}{\hat{U}_\ell}
\newcommand{\al}{\alpha_\ell}

\newcommand{\vac}{\ket{\Phi_L^{\rm vac}}}

\newcommand{\GS}{\ket{\Phi_L^{\rm GS}}}
\newcommand{\EGS}{E_L^{\rm GS}}

\newcommand{\hbS}{\hat{\boldsymbol{S}}}
\newcommand{\hS}{\hat{S}}
\newcommand{\hbr}{\hat{\boldsymbol{r}}}
\newcommand{\hbp}{\hat{\boldsymbol{p}}}

\newcommand{\tJ}{\tilde{J}}

\newcommand{\midskip}{\vspace{.3em}}

\begin{document}

\noindent
{\bf
\Large 
Lieb-Schultz-Mattis theorem with a local twist for general one-dimensional quantum systems
}
\par\bigskip

\noindent
Hal Tasaki\footnote{
Department of Physics, Gakushuin University, Mejiro, Toshima-ku, 
Tokyo 171-8588, Japan
}
\begin{quotation}
We formulate and prove the local twist version of the Yamanaka-Oshikawa-Affleck theorem, an extension of the Lieb-Schultz-Mattis theorem, for one-dimensional systems of quantum particles or spins.
We can treat almost any translationally invariant system wth global $U(1)$ symmetry.
Time-reversal or inversion symmetry is not assumed.
It is proved that, when the ``filling factor'' is not an integer, a ground state without any long-range order must be accompanied by  low-lying excitations whose number grows indefinitely as the system size is increased.
The result is closely related to the absence of topological order in one-dimension.

The present paper is written in a self-contained manner, and does not require any knowledge of the Lieb-Schultz-Mattis and related theorems.
\end{quotation}
\tableofcontents
\section{Introduction}

The Lieb-Schultz-Mattis theorem \cite{LSM}, along with its various extensions \cite{AL,OYA,YOA,Koma,NMT,O,H1,H2,NS,PTAV,WPVZ}, is one of few general arguments for quantum many-body systems which apply to a wide class of models and lead to quantitative results.
In the present paper we study general quantum systems on the one-dimensional lattice, including both particle systems and spin systems, which have non-integral ``filling factor'', and prove the local twist version of the Lieb-Schultz-Mattis theorem.
The only essential requirements are the translation invariance (with period $p$) and the presence of global $U(1)$ symmetry, i.e., the particle number conservation law in particle systems and the invariance under the rotation around a single spin axis (the z-axis) for spin systems.
Time-reversal or inversion symmetry is not assumed.
This completes the project initiated by Oshikawa, Yamanaka, and Affleck \cite{OYA,YOA}, who first extended the theorem to one-dimensional quantum systems with general filling factor.

As in the previous applications of the Lieb-Schultz-Mattis and related theorems, we show that a translation invariant ground state without any long-range order must be accompanied by a low-energy eigenstate.
As a bonus of proving the local twist version of the theorem, we are also able to show that the number of such low-energy excitations grows indefinitely as the system size is increased.\footnote{
As far as we know this implication of the local version of the Lieb-Schultz-Mattis theorem has not been pointed out before.
A closely related observation for infinite systems was made by Koma \cite{Koma}.
}
This in particular means that, in a one-dimensional system with a non-integral filling, {\em a disordered gapped ground state must be infinitely degenerate}\/.
This rigorous result is closely related to the absence of topological order in one-dimension.

Our result was briefly announced in \cite{YOA}, and partially described in an unpublished note \cite{T}.
We shall here present a complete proof including that of a new result for the number of excitations (Corollary~2 and Theorem~2).

\bigskip
Before going into details of our theory, it may be useful to describe some background about the Lieb-Schultz-Mattis theorem and its extensions.

{\em Lieb-Schultz-Mattis:}\/
The original Lieb-Schultz-Mattis theorem, Theorem~2 in Appendix~B of \cite{LSM}, was applied to the one-dimensional Heisenberg antiferromagnet with $S=1/2$, which is one of the most well studied models of quantum magnets.
By using a variational argument, it was proved that the model has an excited state whose excitation energy is bounded from above by a constant times $L^{-1}$, where $L$ is the system size.
This fact suggests (but does not yet prove) that the model has a continuum of gapless excitations in the infinite volume limit.

\begin{figure}
\centerline{\includegraphics[width=12cm,clip]{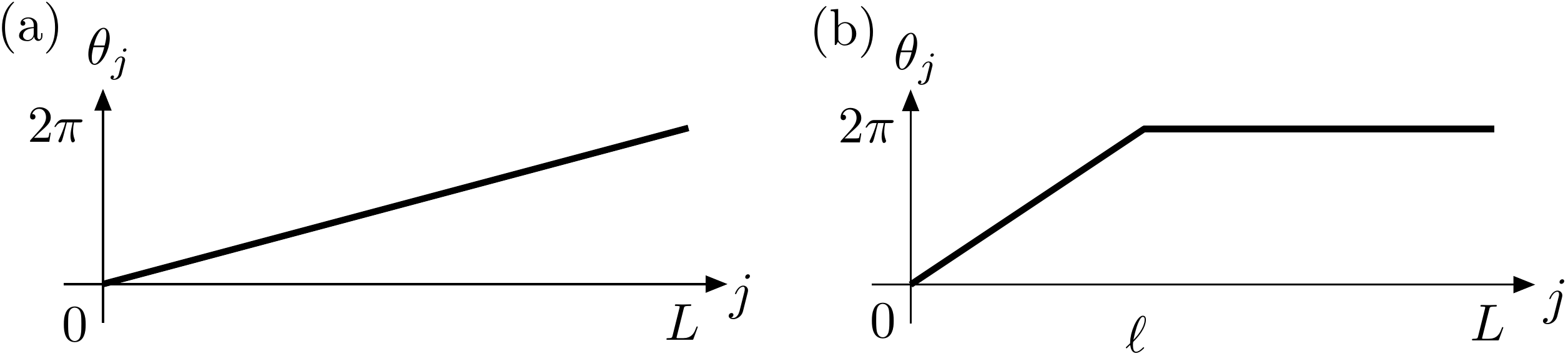}}
\caption[dummy]{
The twist operator $\exp[i\sum_{j=1}^L\theta_j\hS^{(\mrz)}_j]$ is determined by site dependent rotational angle $\theta_j$.
(a)~In the global twist used in the original Lieb-Schultz-Mattis paper, $\theta_j$ varies slowly from 0 to $2\pi$ over the whole lattice, modifying the state in a global manner.
(b)~In the local twist introduced by Affleck and Lieb, $\theta_j$ varies slowly from 0 to $2\pi$ in a finite interval of length $\ell$, where one usually sets $\ell\ll L$.
The local twist operator acts on the ground state locally.
}
\label{f:twist}
\end{figure}

An interesting and essential point of the variational argument of Lieb, Schultz, and Mattis is that it does not rely on explicit forms of the ground state or trial states.
Roughly speaking one only needs to know that the ground state $\GS$ of the system with size $L$ is invariant under the uniform rotation around the z-axix, i.e., $\exp[i\theta\sum_{j=1}^L\hS^{(\mrz)}_j]\GS=\GS$.
One then considers a rotation in which the rotational angle $\theta_j$ varies very slowly from 0 to $2\pi$ (where $2\pi$ is of course equivalent to 0) over the whole lattice as in Figure~\ref{f:twist}~(a).
By applying this ``global twist'' to the ground state, one constructs a trial state $\ket{\Psi_L}=\exp[i\sum_{j=1}^L\theta_j\hS^{(\mrz)}_j]\GS$.
Since the global twist modifies the state very slightly, one can easily show that the trial state $\ket{\Psi_L}$ has an energy expectation value only slightly higher than the ground state energy.
By showing that $\ket{\Psi_L}$ is orthogonal to the ground state, one proves the existence of a low-energy eigenstate.

{\em Affleck-Lieb:}\/
Twenty-five years later, triggered by the discovery \cite{Ha1,Ha2} of the Haldane phenomena for integer spin quantum antiferromagnets, Affleck and Lieb \cite{AL} have made two essential refinements of the theory.

First they examined  extensions of the Lieb-Schultz-Mattis theorem to one-dimensional spin systems with arbitrary spin $S=1/2,1,3/2,\ldots$, and found, quite remarkably, that an extension is possible only when $S$ is a half-odd-integer, namely, $S=1/2,3/2,\ldots$.
Here the trial state $\ket{\Psi_L}$ constructed above has low excitation energy for any $S$, but one can prove that  $\ket{\Psi_L}$ and $\GS$ are orthogonal only when $S$ is a half-odd-integer.

Their second refinement is the use of a ``local twist'' in the construction of a trial state.
The original twist operator $\exp[i\sum_{j=1}^L\theta_j\hS^{(\mrz)}_j]$ of Lieb-Schultz-Mattis, where $\theta_j$ varies slowly over the whole lattice, acts globally on the ground state.
In order to discuss properties of the system in the infinite volume limit, it is desirable to  use an operator which acts only locally.
This is realize by a local twist operator in which $\theta_j$ varies only in a finite (but large) part of the lattice as in Figure~\ref{f:twist}~(b).
By using the local twist version of the theorem, it was proved that a one-dimensional spin system with a half-odd-integer $S$ has gapless excitations in the infinite volume limit provided that the ground state is unique.
See also the discussion before Corollary~1b in section~\ref{s:results} for the importance of locality.

{\em Oshikawa-Yamanaka-Affleck and Yamanaka-Oshikawa-Affleck:}\/
Oshikawa, Yamanaka, and Affleck \cite{OYA} made an essential observation that the Lieb-Schultz-Mattis argument can be extended to a spin model in which the ground state has nonvanishing magnetization.
This is relevant to the problem of quantum spin systems under external magnetic field.
They found that the trial state $\ket{\Psi_L}$, constructed in the same manner as the original Lieb-Schultz-Mattis theorem, is orthogonal to the ground state under the assumption that the ``filling factor'' $\nu=(M/L)+S$ is not an integer.
Here the magnetization $M$ is the eigenvalue of $\hS^{(\mrz)}_{\rm tot}=\sum_{j=1}^L\hS^{(\mrz)}_j$ in the ground state, and $S=1/2,1,\ldots$ is the spin quantum number.
For $M=0$, one has $\nu=S$, and recovers the result of Affleck and Lieb \cite{AL}.

Yamanaka, Oshikawa, and Affleck \cite{YOA} further extended their theory to systems of electrons on the one-dimensional lattice.
Now the key quantity $\nu$ can be naturally interpreted as the filling factor of electrons.
We believe that the two papers \cite{OYA,YOA} played essential roles in extending the scope of the Lieb-Schultz-Mattis argument, and making possible many recent applications of the argument to much wider class of systems.

In \cite{OYA} and in the earlier version of \cite{YOA}, only the global twist was used to construct trial states.
This is because the technique used in \cite{AL} to prove the orthogonality of the ground state and the (locally twisted) trial state was not applicable to models with a general non-integral filling factor.
Our contribution in \cite{T} and in the present paper is construction of trial states orthogonal to the ground state in such a situation.

In all the early works \cite{LSM,AL,OYA,YOA}, it was assumed that the system under consideration possesses time-reversal or inversion symmetry.
This assumption was removed by Koma \cite{Koma}, who applied the technique of local twist to quantum Hall systems on a quasi one-dimensional strip.
See also \cite{NMT} for this and other extensions.

{\em Extensions to higher dimensions:}\/
To  properly extend  the Lieb-Schultz-Mattis theorem to systems in two and higher dimensions is quite important, especially because higher dimensional systems may exhibit rich low-energy behavior associated with various topological order \cite{Wen2016ZOO,ZengChenZhouWenBOOK}.
It was known from the beginning however that a naive extension is problematic since a global twist in a $d$-dimensional lattice increases the energy of the ground state by a constant times $L^{d-2}$ where $L$ is the linear size of the system \cite{LSM}.
The energy increase converges to zero as $L\up\infty$ only for $d=1$.

A breakthrough was brought by Oshikawa \cite{O}, who proposed to make use of a combination of a flux insertion and a gauge transformation instead of the twist operation.
He argued that when the filling factor is not an integer, a ground state is either degenerate or accompanied by a gapless excitation.
Hastings \cite{H1,H2} proposed a similar argument for extending the Lieb-Schultz-Mattis theorem to higher dimensions, which was finally refined into a rigorous theorem by Nachtergaele and Sims \cite{NS}.

Although the conclusions of these higher dimensional extensions of the Lieb-Schultz-Mattis theorem are parallel to that of the original one-dimensional theorem, the ideas behind the proofs seem different.
Another essential difference is that there can be no local twist version of the Lieb-Schultz-Mattis argument in two or higher dimensions.
By a local twist we mean an operation which acts only on a finite subregion (e.g., $\ell\times\cdots\times\ell$ box) of the lattice.
This point will be discussed at the end of section~\ref{s:results}.

For recent (not yet completely rigorous) refinements of the Lieb-Schultz-Mattis argument (or, more precisely, Oshikawa's argument) which take into account the space group symmetry in higher dimensional systems, see, e.g., \cite{PTAV,WPVZ}.

\section{Systems of spinless particles on the one-dimensional lattice}
\label{s:main}

In the present section we shall describe our assumptions (section~\ref{s:setting}) and results (section~\ref{s:results}) carefully in the setting of spinless particles on the one-dimensional lattice.
We believe that one can see the essence of the theory in this simple setting.
We also discuss some trivial examples in section~\ref{s:ex}.

Proofs are given separately in section~\ref{s:proof}.
All the results are easily extended to more realistic systems as we shall see in section~\ref{s:extensions}.
We discuss extensions to a general class of quantum spin chains invariant under rotation about a single axis, a general class of lattice electron systems including the Hubbard model, and a class of tight-binding electron systems in which the positions of the lattice sites are also treated as quantum mechanical degree of freedom.


\subsection{Setting}
\label{s:setting}
We consider a one-dimensional lattice with $pL$ sites, where the fixed positive integer $p$ represents the period (or the number of sites in the unit cell), and $L$ is a sufficiently large positive integer.
We denote the lattice sites as $j,k=1,\ldots,pL$, and use periodic boundary conditions to identify $j$ with $j+pL$.
With each site $j$, we associate operators $\hc_j$ and $\hcd_j$ which annihilates and creates, respectively, a particle at site $j$.
We can treat either fermions with canonical anticommutation relations $\{\hc_j,\hc_k\}=0$, $\{\hc_j,\hcd_k\}=\delta_{j,k}$ or bosons with canonical commutation relations $[\hc_j,\hc_k]=0$, $[\hc_j,\hcd_k]=\delta_{j,k}$.
As usual $\hn_j:=\hcd_j\hc_j$ is the number operator at site $j$.
Let $\hT$ be the operator which generates the translation $j\to j+p$.
It thus satisfies, e.g.,
\eq
\hTd\hc_j\hT=\hc_{j-p}.
\lb{Tdef}
\en

We consider a general local Hamiltonian which conserves the total number of particles, and is invariant under translation by $p$.
More precisely we define
\eq
\HL=-\sum_{j,k=1}^{pL}t_{j,k}\,\hcd_j\hc_k+\sum_{j=1}^{pL}\hV_j,
\lb{HL}
\en
where the first term represents particle hopping and the second term the interaction.
The hopping amplitude $t_{j,k}$ satisfies\footnote{
In \cite{YOA} the hopping amplitude was assumed to be real to ensure time-reversal symmetry.
This condition was removed in \cite{Koma}.
}
  $t_{j,k}=(t_{k,j})^*\in\bbC$.
They are periodic $t_{j,k}=t_{j+p,k+p}$, and short ranged, i.e., $t_{j,k}=0$ whenever $j=k$ or $|j-k|>r$ where the range $r$ is a fixed constant.
We define
\eq
\bar{t}:=\max_j\sum_{k=1}^{pL}|t_{j,k}|,
\lb{tbar}
\en
which is a finite quantity that characterizes the magnitude of the hopping.
The interaction $\hV_j$ is an arbitrary Hermitian operator which depends only on $\hn_j,\hn_{j+1},\ldots,\hn_{j+r}$.
The simplest examples are $\hV_j=v_j^{(1)}\hn_j+v_j^{(2)}\hn_j(\hn_j-1)$ for bosons (where the first term is the on-site potential and the second term is the on-site two-body interaction) and $\hV_j=v_j^{(1)}\hn_j+v_j^{(2)}\hn_j\hn_{j+1}$ for fermions (where the second term is the nearest-neighbor interaction).
We assume the periodicity $\hTd\hV_j\hT=\hV_{j-p}$.
The total Hamiltonian then becomes translation invariant, $\hTd\HL\hT=\HL$.

\begin{figure}
\centerline{\includegraphics[width=6cm,clip]{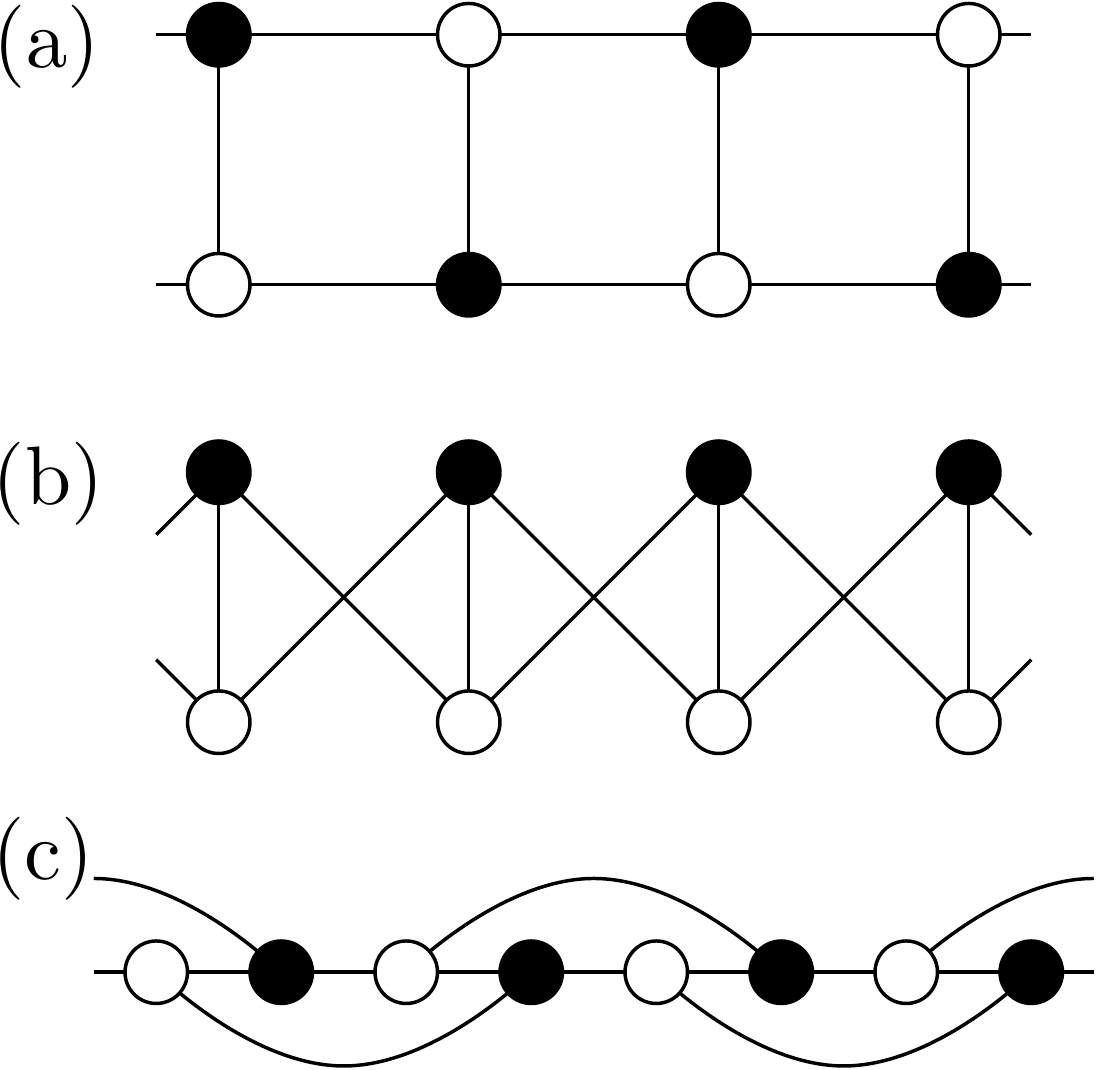}}
\caption[dummy]{
(a)~A model on the two-legged ladder with two kinds of sites.  
Black and white sites are distinguished by, e.g., on-site potentials.
In the original form (a), there are four sites in the unit cell (in the crystalographic sense), but by partially twisting the ladder as in (b), one can rewrite the model into the one-dimensional model in (c) which has period $p=2$.
}
\label{f:ladder}
\end{figure}

Note that various quasi one-dimensional models defined, e.g., on a ladder can be written in the present form.
See Figure~\ref{f:ladder} for an example.

Let the filling factor (per unit cell) $\nu$  be an arbitrary real quantity with $0<\nu<p$ for fermions and $\nu>0$ for bosons.
We set the particle number as $N=[\nu L]$, where $[x]$ denotes the largest integer which does not exceed $x$.
We consider the Hilbert space with exactly $N$ particles on the lattice.

In what follows, we shall fix $t_{j,k}$, $\hV_j$, $p$, $r$, and $\nu$, and let $L$ and $N$ become macroscopically large.

Let $\GS$ be a ground state of $\HL$ which is translation invariant, i.e.,
\eq
\hT\GS=e^{i\kappa}\GS,
\lb{TGS}
\en
for some $\kappa\in\bbR$.
One can always find a ground state satisfying \rlb{TGS} since the Hamiltonian is translation invariant.
When the ground state (for finite $L$) is unique \rlb{TGS} is automatically satisfied.
We denote by $\EGS$ the ground state energy.

\subsection{Main results}
\label{s:results}
Let us first recall the theorem by Yamanaka, Oshikawa, and Affleck \cite{YOA}, in which the original Lieb-Schultz-Mattis theorem was extended to the class of models described in the previous section (and also those treated in section~\ref{s:extensions}).
It will turn out that this is a special case of Corollary~1a below.

\bigno
{\bf Theorem 0:} (Yamanaka, Oshikawa, and Affleck \cite{YOA}, Koma \cite{Koma})
There is a unitary operator $\hat{U}_L$, which is obtained by setting $\ell=L$ in \rlb{Ul}, such that $\bkGS{\hat{U}_L^\dagger\HL\hat{U}_L}-\EGS\le C/L$, where $C$ is a constant given by \rlb{C} below.
When the filling factor $\nu$ is not an integer, one further has $\bkGS{\hat{U}_L}=0$, and hence the state $\hat{U}_L\GS$ is orthogonal to $\GS$.
This means that there is an energy eigenstate orthogonal to $\GS$ and whose energy eigenvalue $E$ satisfies $E-\EGS\le C/L$.

\bigskip

This theorem makes use of the ``global twist'' applied to the ground state.
Let us describe our results based on the ``local twist''.

For $\ell=1,2,\ldots,L$, we define the density operator
\eq
\rl:=\frac{1}{\ell}\sum_{j=1}^{p\ell}\hn_j.
\lb{rho}
\en
Note that the translation invariance implies
\eq
\bkGS{\rl}=\frac{N}{L}=\nu+O(L^{-1}).
\lb{rlNL}
\en
We also define
\eq
\Drl:=\rl-\nu.
\lb{Drho}
\en

Then our first result is the following.
It follows from a more mathematical statement summarized as Theorem~1 in section~\ref{s:proof}.
\bigno
{\bf Corollary 1a:} Suppose that the filling factor $\nu$ is not an integer, and the variance $\bkGS{(\Drl)^2}$ can be made as small as one wishes by letting $\ell$ large (where $L$ is assumed to be much larger than $\ell$).
Then, for sufficiently large $\ell$, there exists a local operator $\Wl$ which depends only on $\hn_j$ with $j=1,\ldots,p\ell$, and satisfies
\eqg
\bkGS{\Wl}=0,
\lb{GWG}\\
\bkGS{\Wl^\dagger\Wl}=1,
\lb{GWWG}
\eng
and
\eq
\bkGS{\Wl^\dagger\HL\Wl}-\EGS\le\frac{C}{\ell},
\lb{VE}
\en
where the ($L$ independent) constant $C$ is given by
\eq
C=\frac{8\pi^2(p+1)(r+1)^2}{p^2}\nu\bar{t}.
\lb{C}
\en

\bigskip
We first check that Theorem~0 is a special case of Corollary~1a.
If we set $\ell$ equal to $L$, then the operator $\Wl$ coincides with the global twist operator $\hat{U}_L$ in Theorem~0.
Moreover the operator $\hat{\rho}_L$ is equal to the constant $N/L=\nu+O(L^{-1})$ in our Hilbert space, and hence we always have $\bkGS{(\Di\hat{\rho}_L)^2}=O(L^{-2})$;
the condition about the variance in the corollary is automatically  satisfied.

We next consider the opposite case with $\ell\ll L$.
Since $\bkGS{(\Drl)^2}$ is the variance\footnote{
To be very precise, this quantity may be slightly different from the variance since $\nu$ may differ from $\bkGS{\rl}$ by $O(1/L)$ as in \rlb{rlNL}.
But this difference is negligible for large $L$.
} of the density within the interval $[1,p\ell]$, it usually decreases toward zero as $\ell$ grows.
More specifically, $\bkGS{(\Drl)^2}$ decays to zero as $\ell\up\infty$ if the truncated correlation function $\bkGS{\hn_j\hn_k}-\bkGS{\hn_j}\bkGS{\hn_k}$ decays to zero as $|j-k|\up\infty$ (where $L$ is supposed to grow properly).
Note that truncated correlation functions fail to decay only when the state has long-range order.
To sum the condition about the variance in Corollary 1a is satisfied if the ground state does not exhibit any long-range order.

\bigno{\em Remarks:}\/
1.~More precisely truncated correlation functions fail to decay when the ground state has long-range order but does not exhibit corresponding spontaneous symmetry breaking.
In such a situation there exist more than one infinite volume ground states.
See example~4 of section~\ref{s:ex} for a simple example and  \cite{KT} for a general theory.
\par\midskip\noindent
2.~If $\bkGS{(\Drl)^2}\simeq(\text{constant})\ne0$, and the assumption for Corollary~1a is not satisfied, one can follow Watanabe \cite{W} to show that\footnote{
To be precise, one lets $\Drl':=\rl-\bkGS{\rl}=\Drl+O(1/L)$, and considers the trial state $\ket{\Xi}=\Drl'\GS/\sqrt{\bkGS{(\Drl')^2}}$, which obviously satisfies $\langle\Phi^{\rm GS}_L\ket{\Xi}=0$, and also satisfies $\sbkt{\Gamma|\HL|\Gamma}-\EGS\le(\text{const})\times\ell^{-2}$.
To show the latter variational estimate, we use the identity $\sbkt{\Xi|\HL|\Xi}-\EGS=(1/2)\bkGS{[[\Drl',\HL],\Drl']}/\bkGS{(\Drl')^2}$, and note that $\Vert [[\Drl',\HL],\Drl']\Vert=O(\ell^{-2})$.
}   the state $\Drl\GS$ is orthogonal to the ground state and has excitation energy not greater than $(\text{const})\times\ell^{-2}$.
Then the standard variational argument implies that there exits an energy eigenstate orthogonal to the ground state whose energy eigenvalue $E$ satisfies $E-\EGS\le(\text{const})\times\ell^{-2}$.
Note that Theorem~0 applies to this case as well.  In this sense Theorem~0 covers different situations with essentially different natures of low energy excitations.

\bigskip

Now let us assume that the condition for the variance $\bkGS{(\Drl)^2}$, as well as that for $\nu$, in Corollary~1a is satisfied.
This is true if  the ground state does not exhibit any long-range order.
Then \rlb{GWG}, \rlb{GWWG}, and \rlb{VE} imply that $\Wl\GS$ is a normalized state orthogonal to the ground state $\GS$ whose energy expectation value does not exceed $\EGS+(C/\ell)$.
Then the standard variational argument implies that there exists an energy eigenstate $\ket{\Psi}$ orthogonal to $\GS$ whose energy eigenvalue $E'$ satisfies $E'-\EGS\le C/\ell$.
Note that $\ell$ can be made as large as one wishes (provided that $L$ is large).
This proves the existence of a low-energy eigenstate.

We stress that the trial state $\Wl\GS$ is obtained by applying the local twist operator  $\Wl$ onto the ground state $\GS$.
The locality may be natural from an experimental point of view since one normally excites a ground state by using some local operation.\footnote{
Of course $\GS$ and $\Wl\GS$ may differ globally even when $\Wl$ is local.
See example~2 in section~\ref{s:ex}.
}
However the main motivation for considering local twist in \cite{AL} comes from the treatment of the infinite volume limit.
When one lets $L\up\infty$, it may happen that the ground state $\GS$ and its global twist  $\hat{U}_L\GS$ converge to a single infinite volume state.
In this case one does not obtain any information about excitations (in the infinite volume limit) from $\hat{U}_L\GS$.
If one uses the local twist operator $\Wl$ with a large but fixed $\ell$, then it is guaranteed that $\GS$ and $\Wl\GS$ tend to different infinite volume states as $L\up\infty$.

By following Affleck and Lieb \cite{AL} one can make this observation into a mathematical argument to prove the following from Theorem~1.
See  \cite{AL} (and also \cite{KT}) for precise formulation of infinite systems.

\bigno
{\bf Corollary 1b:}
When $\nu$ is not an integer, the infinite volume limit of the model has either (a)~more than one ground states or (b)~a unique ground state with gapless excitations. 

\bigskip

Here we shall stick onto finite systems, and see another implication of the results based on the local twist.

Take some positive integer $s$, and consider the operator $\hT^s\Wl(\hTd)^s$, which is 
the translation of $\Wl$ by a distance $sp$.
It is expected in many cases that the trial states $\Wl\GS$ and $\hT^s\Wl(\hTd)^s\GS$ are distinct for sufficiently large $s$.
This suggests that one can construct multiple trial states which are linearly independent and also have low excitation energies.

The following is the rigorous result in this direction, which follows from Theorem~2 stated and proved in section~\ref{s:proof}.
\bigno
{\bf Corollary~2:}
Suppose that the filling factor $\nu$ is not an integer and the expectation value
$\bkGS{(\Wl)^\dagger\,\{\hT^s\Wl(\hTd)^s\}}$ can be made as small as one wishes by letting $s$ large (where $L$ is assumed to be sufficiently large).
Take any (small) $\lambda>0$ and a (large) positive integer $n$.
Then one can find a sufficiently large $L$ such that there are $n$ eigenstates $\ket{\Psi_L^{(1)}},\ldots,\ket{\Psi_L^{(n)}}$ of $\HL$ which are mutually orthogonal and orthogonal to $\GS$.
The energy eigenvalue $E_L^{(\mu)}$ of $\ket{\Psi_L^{(\mu)}}$ satisfies $E_L^{(\mu)}-\EGS\le\lambda$ for any  $\mu=1,\ldots,n$.
\bigskip

Note that $\bkGS{(\Wl)^\dagger}\bkGS{\hT^s\Wl(\hTd)^s}=\bkGS{(\Wl)^\dagger}\bkGS{\Wl}=0$ by translation invariance and  \rlb{GWG}.
Thus the second assumption is again about the decay of the truncated correlation function 
\eq
\bkGS{(\Wl)^\dagger\,\{\hT^s\Wl(\hTd)^s\}}-\bkGS{(\Wl)^\dagger}\bkGS{\hT^s\Wl(\hTd)^s}
\en
of complicated local operators $(\Wl)^\dagger$ and $\hT^s\Wl(\hTd)^s$.
Thus the assumption is valid if the ground state does not exhibit any long-range order.

To summarize we have proved that, in a one-dimensional system (which belongs to the class we consider) with a non-integral filling factor $\nu$, one has at least one of the following two alternatives.
\bigno
(A)~The ground state exhibits a certain long-range order.
\par\noindent
(B)~There exist low-energy eigenstates whose number increases indefinitely as $L$ grows.

\bigskip
A gapless system which possesses a continuum of excitations directly above the ground state falls into the case (B).

To see the implications on a gapfull system, assume that there is a model which has ground states and near ground states with almost degenerate energies, separated by a finite gap of order 1 from other energy eigenstates.
More precisely, we assume that there is very small $\ep>0$ and a finite gap $\gamma>0$, and there are no energy eigenvalues in the interval $(\EGS+\ep,\EGS+\gamma)$.
By using Corollary~2 by setting $\lambda\lesssim\gamma$, we see that all the $n$ energy eigenstates $\ket{\Psi_L^{(1)}},\ldots,\ket{\Psi_L^{(n)}}$ have energy eigenvalues in the interval $[\EGS,\EGS+\ep]$ and hence are near ground states.
We thus conclude that the degeneracy of near ground states must grow indefinitely with the system size $L$.

We can thus refine the alternative (B) above as follows.

\bigno
{\bf Conclusion:}
In a one-dimensional system (which belongs to the class we consider) with a non-integral filling factor $\nu$, one has at least one of the following three possibilities.
\par\noindent
\ (A)~The ground state exhibits certain long-range order.
\par\noindent
\ (B1)~The system is gapless and there exist low-energy eigenstates whose number increases indefinitely as $L$ grows.
\par\noindent
\ (B2)~The system is gapfull and the ground state degeneracy (more precisely, the number of near ground states) increases indefinitely as $L$ grows.
\par
In particular we have completely ruled out the possibility
\par\noindent
\ (C)~There are a finite number\footnote{
More precisely we mean that the number of near ground states are bounded when $L$ grows.
} of (near) ground states which do not have long-range order and are separated from other energy eigenstates by a nonvanishing gap.

\bigskip

There are many examples of systems with non-integral $\nu$ which show behaviors corresponding to (A), (B1), or (B2).
The case (A) is realized, e.g., in a system with crystalline ordering (see example~4 in section~\ref{s:ex}).
The case (B1) may be most common, and can be seen in free (or weakly interacting) fermion systems (see example~1), or quantum antiferromagnetic chains with half-odd-integer spins.
The case (B2), on the other hand, may be realized in systems with trivial degeneracy such as example~3.

There are, on the other hand, one-dimensional systems with an integer filling $\nu$ which have a unique ground state accompanied by a nonvanishing gap  but do not exhibit  long-range order.  This corresponds to the case (C).
Examples include free fermions (see example~1) and quantum antiferromagnetic chains with integer spins.
In this sense the Lieb-Schultz-Mattis argument sharply characterizes the difference between systems with integral and non-integral filling factors.

\bigskip

As we have noted in the introduction, extensions (with various degrees of rigor) of the Lieb-Schultz-Mattis theorem to two and higher dimensions are known \cite{O,H1,H2,NS}.
Roughly speaking one can show a conclusion similar to that of Theorem~0, i.e., a system with a non-integral filling factor inevitably has at least one low-energy eigenstate other than the ground state.
But, in two or higher dimensions, there are no results that correspond to Corollary~2.
We believe that this is inevitable since there are higher dimensional systems with a non-integral filling which are believed to exhibit the behavior (C) above.
For example the fractional quantum Hall system on a torus with $\nu=1/3$ is believed to have three-fold degenerate (or near degenerate) gapped ground states without any long-range order.
This is a typical example of topological order (see, e.g., \cite{Wen2016ZOO,ZengChenZhouWenBOOK,Wen}).
The fact that the case (C) is ruled out in one-dimension is thus closely related to the absence of topological order in one-dimension \cite{CGW,Wen2016ZOO,ZengChenZhouWenBOOK}.\footnote{
Antiferromagnetic quantum spin chain with $S=1$ exhibiting Haldane phenomena is known to possess a kind of hidden order.
But this is related to symmetry protected topological phase, not to topological order.
See, e.g., \cite{Wen2016ZOO,ZengChenZhouWenBOOK}.
}

Recall that Corollary~2 is based on the construction of trial states using the local operator $\Wl$.
If it was possible to prove a local twist version of the Lieb-Schultz-Mattis theorem in two or higher dimensions, it would contradict the above belief about the $\nu=1/3$ fractional quantum Hall effect.
This observation may reflect a fundamental difference  between one and higher dimensions about the structure of  low-energy eigenstates in quantum many-body systems.

\subsection{Simple examples}
\label{s:ex}
We shall discuss  simple examples which illustrate the classification given in Conclusion.
We hope these trivial examples may help the readers grasp the essence of the classification.
We stress, however, that our results are general and apply to nontrivial models whose properties are not easily understood.

\begin{enumerate}
\item {\em Free fermions:}\/
Consider a free fermion system obtained by setting $\hV_j=0$ for all $j$.
The energy eigenstates are fully determined by the band structure that is obtained from the hopping matrix $(t_{j,k})$.
Assume that the $p$ bands have nonvanishing dispersion and are separated by a nonzero (single-particle) energy gap.

When $\nu$ is an integer, the lowest $\nu$ bands are fully filled in the ground state, and there is a nonvanishing gap.
This corresponds to the case (C).

When $\nu$ is not an integer, then one of the bands is partially filled in the ground state, and there is a continuum of gapless excitations.
If one turns on sufficiently weak short-range interaction, the low-energy behavior of the system is likely to be described as a Tomonaga-Luttinger liquid.
Hence the ground state remains to be unique and gapless.
These are examples of (B1).

\item {\em Fermions with nearest-neighbor repulsive interaction at $\nu=1/2$:}\/
This is a toy model of Mott insulator.
Consider a system of fermions with $t_{j,k}=0$ for any $j,k$, and $\hV_j=v_0\hn_j\hn_{j+1}$ where $v_0>0$.
The model has period $p=1$.
Assume that $L$ is even, and set $N=L/2$ to have $\nu=1/2$.
Clearly there are two ground states $\ket{\Phi_L^{\rm GS, even}}=(\prod_{j=1}^{L/2}\hcd_{2j})\vac$ and $\ket{\Phi_L^{\rm GS, odd}}=(\prod_{j=1}^{L/2}\hcd_{2j-1})\vac$, with $\EGS=0$, and there is an excitation gap $v_0>0$.
(Here $\vac$ is the state with no particles on the chain.)
This may seem to be the case (C), in apparent contradiction with our conclusion.
But this example should be classified as (A), i.e., there is long-range order.

To see this, take two translation invariant ground states\footnote{
If one adds a small hopping $t_{j,j+1}=t\ne0$, then one of $\ket{\Phi_L^{\rm GS,\pm}}$ becomes the unique ground state and the other becomes the near ground state with almost degenerate energy.
} $\ket{\Phi_L^{\rm GS,\pm}}:=\{\ket{\Phi_L^{\rm GS, even}}\pm\ket{\Phi_L^{\rm GS, odd}}\}/\sqrt{2}$, and denote the corresponding expectation values as $\sbkt{\cdots}^\pm=\sbkt{\Phi_L^{\rm GS,\pm}|\cdots|\Phi_L^{\rm GS,\pm}}$.
One easily finds that $\sbkt{\hn_j}^\pm=1/2$ for any $j$, while $\sbkt{\hn_j\hn_k}^\pm=1/2$ if $j-k\ne0$ is even and $\sbkt{\hn_j\hn_k}^\pm=0$ if $j-k$ is odd.
Thus the truncated correlation function $\sbkt{\hn_j\hn_k}^\pm-\sbkt{\hn_j}^\pm\sbkt{\hn_k}^\pm$ oscillates between $1/4$ and $-1/4$, and never decays.
This is an indication of long-range order without symmetry breaking.

Let us note in passing that in this model one has $\sbkt{\Drl}=0$ in any of the above ground states provided that $\ell$ is even.
This means that Corollary~1a (not Corollary~2) happened to be valid in this model, in spite of the presence of the long-range order.
It is indeed not hard to see that $\ket{\Phi_L^{\rm GS,\mp}}\propto\Wl\ket{\Phi_L^{\rm GS,\pm}}$.
Note also that, in this case, the application of the local operator $\Wl$ modifies the ground states $\ket{\Phi_L^{\rm GS,\pm}}$ in a global manner.

\item {\em Fermions with nearest-neighbor repulsive interaction at $\nu<1/2$:}\/
Take the same model but suppose that $N/L=\nu<1/2$.
Then any particle configuration in which two particles do not occupy neighboring sites gives a ground state, and there is a gap $v_0>0$.
Since the ground state degeneracy grows indefinitely as $L$ grows, one sees that this is an example of (B2).

\item {\em Fermions with nearest-neighbor attractive interaction:}\/
This is a toy model example in which the assumption of Corollary~1 does not hold.
Set $t_{j,k}=0$ for any $j,k$, and $\hV_j=-v_0\hn_j\hn_{j+1}$ where $v_0>0$.
Suppose that $N/L=\nu<1$.
Then there are $L$-fold degenerate ground states given by $\ket{\Phi_L^{{\rm GS},j}}:=(\prod_{k=j}^{j+N-1}\hcd_k)\vac$ with $j=1,\ldots,L$, in which $N$ particles are forming a single cluster.
Note that these ground states spontaneously break the translation symmetry.

A translation invariant ground state can be formed as $\GS=L^{-1/2}\sum_{j=1}^L\ket{\Phi_L^{{\rm GS},j}}$.
An inspection shows that, in this ground state, the density $\rl$ is almost 1 with probability $\nu$ and is almost vanishing with probability $1-\nu$, provided that $\ell\ll L$.
One finds that $\bkGS{(\Drl)^2}\simeq\nu$, and hence the condition for Corollary~1 is never satisfied no matter how large $\ell$ is.
\end{enumerate}

\section{Proof}
\label{s:proof}
We shall prove all the results within the setting of spinless particles.
We stress that all argument here can be easily extended to general models treated in section~\ref{s:extensions}.

For any positive integer $\ell$ with $\ell\le L$, we define an unitary operator by
\eq
\Ul:=\exp\bigl[i\sum_{j=1}^{p\ell}\theta_j\,\hn_j\bigr],
\lb{Ul}
\en
with
\eq
\theta_j=
\begin{cases}\frac{2\pi}{\ell}\Bigl[\frac{j-1}{p}\Bigr]&j=1,\ldots,p\ell;\\
2\pi&j=p\ell+1,\ldots,pL,\\
\end{cases}
\lb{tj}
\en
where $[x]$ in \rlb{tj} is the largest integer that does not exceed $x$.
See Figure~\ref{f:twist}~(b).
We have defined $\theta_j$ for $j>p\ell$ for later convenience.
As we noted earlier $\hat{U}_L$ is the global twist operator constructed originally in \cite{LSM}.
When $\ell\ll L$, the operator $\Ul$ describes a ``local twist'' in the range $1\le j\le p\ell$.

Note that for any $\theta\in\bbR$, one trivially has
\eq
\exp\bigl[-i\sum_{j=1}^{pL}\theta\,\hn_j\bigr]\,\HL\,\exp\bigl[i\sum_{j=1}^{pL}\theta\,\hn_j\bigr]=\HL,
\en
because $\sum_{j=1}^{pL}\hn_j$ is always equal to $N$ in our Hilbert space.
This is the global $U(1)$ invariance, which is essential to the Lieb-Schultz-Mattis argument.
Since $\theta_j$ of \rlb{tj} varies slowly with $j$, it is expected that the unitary operator $\Ul$ changes $\HL$ only slightly.  See \rlb{UHUH} below.
This observation leads to the following lemma \cite{YOA,Koma} which shows that the twisted state $\Ul\GS$ has a small excitation energy.

\bigno
{\bf Lemma~1:}
For any $\ell$ and $L$ such that $\max\{1,2r-p\}\le\ell\le L$,  one has
\eq
\DE:=\bkGS{\Ul^\dagger\HL\Ul}-\EGS\le\frac{C}{2\ell},
\lb{DE}
\en
where the ($L$ independent) constant $C$ is given in \rlb{C}.

\bigno
{\em Proof:}\/
By using the easily verifiable relations
\eq
e^{-i\theta\hn_j}\hc_je^{i\theta\hn_j}=e^{i\theta}\hc_j,\quad
e^{-i\theta\hn_j}\hcd_je^{i\theta\hn_j}=e^{-i\theta}\hcd_j,\quad
e^{-i\theta\hn_j}\hn_je^{i\theta\hn_j}=\hn_j,
\lb{eceene}
\en
one finds that
\eq
\Ul^\dagger\HL\Ul-\HL
=\sum_{j,k=1}^{pL}(1-e^{-i(\theta_j-\theta_k)})t_{j,k}\hcd_j\hc_k.
\lb{UHUH}
\en
Note that we have extended the sum to that over the whole lattice (recall that $\theta_j=2\pi$ for $j>p\ell$).
By replacing $\theta_j$ with $-\theta_j$, we also have
\eq
\Ul\HL\Ul^\dagger-\HL
=\sum_{j,k=1}^{pL}(1-e^{i(\theta_j-\theta_k)})t_{j,k}\hcd_j\hc_k.
\lb{UHUH2}
\en

Let us abbreviate ground state expectation values $\bkGS{\cdots}$ as $\sbkt{\cdots}$.
Then the energy difference defined in \rlb{DE} is bounded as
\eqa
\DE&=\sbkt{\Ul^\dagger\HL\Ul-\HL}
\nl&\le\sbkt{\Ul^\dagger\HL\Ul-\HL}+\sbkt{\Ul\HL\Ul^\dagger-\HL}
\nl&
=\sum_{j,k=1}^{pL}2\bigr\{1-\cos(\theta_j-\theta_k)\bigl\}t_{j,k}\sbkt{\hcd_j\hc_k},
\lb{DE2}
\ena
where the second line follows from the trivial inequality\footnote{
This seemingly trivial observation in \cite{Koma} was essential in removing the assumption about time-reversal (or inversion) symmetry.
} $\bkGS{\Ul\HL\Ul^\dagger}\ge\EGS=\bkGS{\HL}$, and we used \rlb{UHUH} and \rlb{UHUH2} to get the final expression.

We can bound the correlation function in the right-hand side of \rlb{DE2} by using the Schwarz inequality\footnote{
\label{fn:Sch}
For any operators $\hat{A}$, $\hat{B}$, one has $\bigl|\sbkt{\hat{A}^\dagger\hat{B}}\bigr|^2\le\sbkt{\hat{A}^\dagger\hat{A}}\sbkt{\hat{B}^\dagger\hat{B}}$, where $\sbkt{\cdots}$ is any expectation value.
} as
\eq
\sbkt{\hcd_j\hc_k}\le\sqrt{\sbkt{\hn_j}\sbkt{\hn_k}}
\le\max_j\sbkt{\hn_j}
\le\sum_{j=1}^p\sbkt{\hn_j}=\frac{N}{L}\le\nu.
\lb{ccnu}
\en
The first factor in the right-hand side of \rlb{DE2} is bounded by using $\cos x\ge1-x^2/2$ as
\eq
2\bigr\{1-\cos(\theta_j-\theta_k)\bigl\}\le(\Di\theta_{j,k})^2,
\en
with $\Di\theta_{j,k}=\theta_j-\theta_k\ (\mathrm{mod}\,2\pi)$ is chosen to satisfy $|\Di\theta_{j,k}|<2\pi$.
Recall that $t_{j,k}=0$ if $|j-k|>r$.
This means that $\Di\theta_{j,k}$ contributing to the sum in the right-hand side of \rlb{DE2} always satisfies
\eq
|\Di\theta_{j,k}|\le\frac{2\pi}{\ell p}(r+1).
\en
Noting that $\theta_j=0 \ (\mathrm{mod}\,2\pi)$ unless $p+1\le j\le \ell p$, we see that $\Di\theta_{j,k}$ can be nonvanishing only when $p+1\le j\le \ell p$ or $p+1\le k\le\ell p$.
Putting all these estimates together we get
\eqa
\DE&\le\biggl\{\frac{2\pi}{p \ell}(r+1)\biggr\}^2\,\nu\sum_{j,k=p+1-r}^{\ell p+r}|t_{j,k}|
\nl
&\le\biggl\{\frac{2\pi}{p \ell}(r+1)\biggr\}^2\,\nu\,\{p(\ell-1)+2r\}\,\bar{t},
\ena
where $\bar{t}$ is defined in \rlb{tbar}.
Finally to make the expression simpler we note that $p(\ell-1)+2r\le(p+1)\ell$ if $\ell\ge 2r-p$, and find
\eq
\DE\le\biggl\{\frac{2\pi}{p \ell}(r+1)\biggr\}^2\,\nu(p+1)\ell\bar{t}
=\frac{4\pi^2(p+1)(r+1)^2}{p^2}\nu\bar{t}\times\frac{1}{\ell},
\en
which is the desired \rlb{DE}.~\qedm

\bigskip

The variational estimate \rlb{DE} may seem to imply that $\Ul\GS$ is an excited state with excitation energy smaller than $C/(2\ell)$.
This is indeed not the case since  $\Ul\GS$ is in general not orthogonal to $\GS$.
We can however prove ``near orthogonality'' provided that the filling factor $\nu$ is not an integer and the variance of $\Di\rl$ defined in \rlb{Drho} is sufficiently small.
This idea is due to Oshikawa \cite{Opc}.
The main key for a rigorous proof is the following inequality derived by Koma (section~5 of \cite{Koma}), who improved our  earlier estimate in \cite{T}.

\bigno
{\bf Lemma 2:}
If $\nu$ is not an integer, we have
\eq
\bigl|\bkGS{\Ul}\bigr|\le
\frac{\pi^2\,\bkGS{(\Di\rl)^2}}{(\sin \pi\nu)^2}.
\lb{dr2}
\en
\bigno
{\em Proof:}\/
Since \rlb{tj} implies $\theta_j=0$ if $1\le j\le p$ and $\theta_j=2\pi$ if $\ell p+1\le j\le(\ell+1)p$, we can write $\Ul$ (which is defined in \rlb{Ul}) as
\eq
\Ul=\exp\bigl[i\sum_{j=p+1}^{p(\ell+1)}\theta_j\,\hn_j\bigr].
\en
By using this expression and the definition of translation operator \rlb{Tdef}, we have
\eqa
\hT^\dagger\Ul\hT&=\exp\bigl[i\sum_{j=p+1}^{p(\ell+1)}\theta_j\,\hn_{j-p}\bigr]
=\exp\bigl[i\sum_{j=1}^{p\ell}\theta_{j+p}\,\hn_{j}\bigr]
\nl&=\exp\bigl[\frac{2\pi i}{\ell}\sum_{j=1}^{p\ell}\hn_{j}\bigr]\,\Ul
=\exp[2\pi i \rl]\,\Ul,
\ena
where we noted that $\theta_{j+p}=\theta_j+2\pi/\ell$ if $1\le j\le p\ell$.

By recalling \rlb{TGS}, we have
\eqa
\bkGS{\Ul}&=\bkGS{\hT^\dagger\Ul\hT}=\bkGS{e^{2\pi i\rl}\Ul}
\nl&=e^{2\pi i\nu}\bkGS{\Ul}+\bkGS{(e^{2\pi i\rl}-e^{2\pi i\nu})\Ul}.
\lb{Ueee}
\ena
Let us again write  $\bkGS{\cdots}$ as $\sbkt{\cdots}$.
Then \rlb{Ueee} is rewritten as
\eq
(1-e^{2\pi i\nu})\sbkt{\Ul}=\sbkt{(e^{2\pi i\rl}-e^{2\pi i\nu})\Ul}.
\en
From the Schwarz inequality (see footnote~\ref{fn:Sch}), we then get
\eqa
\bigl|(1-e^{2\pi i\nu})\sbkt{\Ul}\bigr|^2&=\bigl|\sbkt{(e^{2\pi i\rl}-e^{2\pi i\nu})\Ul}\bigr|^2
\nl&\le\sbkt{(e^{2\pi i\rl}-e^{2\pi i\nu})(e^{-2\pi i\rl}-e^{-2\pi i\nu})}\,
\sbkt{\Ul^\dagger\Ul}
\nl&=4\sbkt{\bigl\{\sin[\pi(\rl-\nu)]\bigr\}^2}.
\ena
We thus find from $(\sin x)^2\le x^2$ that
\eq
\bigl|\sbkt{\Ul}\bigr|^2\le\frac{\sbkt{\bigl\{\sin[\pi(\rl-\nu)]\bigr\}^2}}{|\sin\pi\nu|^2}
\le \frac{\pi^2\sbkt{(\rl-\nu)^2}}{|\sin\pi\nu|^2},
\en
which is the desired \rlb{dr2}.~\qedm

\bigskip

From Lemma~1 and Lemma~2, we get our first main result.
Corollary~1 is a straightforward consequence of this theorem.

\bigno
{\bf Theorem~1:}
Suppose that $\nu$ is not an integer, and it holds that
\eq
\bkGS{(\Di\rl)^2}\le\frac{(\sin\pi\nu)^2}{2\pi^2}.
\lb{dr1}
\en
Then the operator $\Wl$ defined by the following \rlb{Wdef} satisfies \rlb{GWG}, \rlb{GWWG}, and \rlb{VE}.

\bigno
{\em Proof:}\/
Let $\al=\bkGS{\Ul}$.
From  \rlb{dr2} and \rlb{dr1}, we have $|\al|^2\le1/2$.
Define
\eq
\Wl:=\frac{\Ul-\al}{\sqrt{1-|\al|^2}},
\lb{Wdef}
\en
which obviously satisfies \rlb{GWG}.
It is also easy to see that
\eq
\bkGS{\Wl^\dagger\Wl}=\frac{\bkGS{(\Ul^\dagger-\al^*)(\Ul-\al)}}{1-|\al|^2}=1,
\en
which is \rlb{GWWG}.
To show the bound \rlb{VE} for the energy expectation value, we note that
\eqa
\bkGS{\Wl^\dagger\HL\Wl}-\EGS&=
\frac{\bkGS{(\Ul^\dagger-\al^*)\HL(\Ul-\al)}-(1-|\al|^2)\EGS}{1-|\al|^2}
\nl
&=\frac{\bkGS{\Ul^\dagger\HL\Ul}-\EGS}{1-|\al|^2}
\le\frac{C}{\ell},
\ena
where we used the variational estimate \rlb{DE} and $|\al|^2\le1/2$.~\qedm

\bigskip

We move onto the second theorem about multiple low energy eigenstates.
Take an arbitrary positive integer $n$, and let $\Wl^{(\mu)}=(\hT)^{s_\mu}\Wl(\hTd)^{s_\mu}$ for $\mu=1,\ldots,n$, where $s_1,\ldots,s_n$ are distinct integers.
Then we have the following.

\bigno
{\bf Theorem 2:}
Suppose that, for some $L$ and some choice of $s_1,\ldots,s_n$, we have
\eq
\bigl|\bkGS{(\Wl^{(\mu)})^\dagger\Wl^{(\zeta)}}\bigr|\le\frac{1}{2n},
\lb{WWcond}
\en
for any $\mu,\zeta=1,\ldots,n$ with $\mu\ne\zeta$.
Then there exist eigenstates $\ket{\Psi_L^{(\mu)}}$ of $\HL$ with energy eigenvalue $E_L^{(\mu)}$ for $\mu=1,\ldots,n$, and we have $\sbkt{\Psi_L^{(\mu)}|\Psi_L^{(\zeta)}}=\delta_{\mu,\zeta}$, $\sbkt{\Psi_L^{(\mu)}|\Phi_L^{\rm GS}}=0$, and
\eq
E_L^{(\mu)}-\EGS\le\frac{2nC}{\ell}.
\en

\bigskip
To prove Corollary~2 (given Theorem~2), we first chose $\ell$ (for given $n$ and $\lambda$) such that $2nC/\ell\le\lambda$.
Then, with this $\ell$, we choose $L$ and $s_1,\ldots,s_n$ which make the condition \rlb{WWcond} valid.  This is always possible by the assumption of Corollary~2.

\bigno
{\em Proof of Theorem~2:}\/
Let $\ket{\Xi^{(\mu)}}=\Wl^{(\mu)}\GS$ for $\mu=1,\ldots,n$, which are normalized  low energy states orthogonal to $\GS$.
We first show that $\ket{\Xi^{(1)}},\ldots,\ket{\Xi^{(n)}}$ are linearly independent.
To see this it suffices to show that the Gramm matrix $(G_{\mu,\zeta})_{\mu,\zeta=1,\ldots,n}$ defined by $G_{\mu,\zeta}=\sbkt{\Xi^{(\mu)}|\Xi^{(\zeta)}}$ is regular.
Take arbitrary $c_1,\ldots,c_n\in\bbC$ such that $\sum_{\mu=1}^n|c_\mu|^2=1$, and note that
\eqa
\sum_{\mu,\zeta=1}^nc_\mu^*G_{\mu,\zeta}c_\zeta&=
\sum_{\mu=1}^n|c_\mu|^2+\sum_{\mu\ne\zeta}c_\mu^*G_{\mu,\zeta}c_\zeta
\ge1-\sum_{\mu\ne\zeta}|c_\mu|\,|G_{\mu,\zeta}|\,|c_\zeta|
\nl
&\ge1-\frac{n-1}{2n}>\frac{1}{2}.
\lb{Gbound}
\ena
Here we noted that $G_{\mu,\mu}=1$ and $|G_{\mu,\zeta}|\le1/(2n)$, which is \rlb{WWcond}, and used the trivial bound
\eq
\sum_{\mu\ne\zeta}|c_\mu|\,|c_\zeta|\le\frac{1}{2}\sum_{\mu\ne\zeta}(|c_\mu|^2+|c_\zeta|^2)=\sum_{\mu\ne\zeta}|c_\mu|^2=(n-1)\sum_{\mu}|c_\mu|^2=n-1.
\lb{cc}
\en
The bound \rlb{Gbound} implies that any eigenvalue of the Gramm matrix $(G_{\mu,\zeta})_{\mu,\zeta=1,\ldots,n}$ is greater than $1/2$.
This proves the desired regularity and hence the linear independence.

We thus see that the states $\ket{\Xi^{(1)}},\ldots,\ket{\Xi^{(n)}}$ span $n$ dimensional subspace orthogonal to $\GS$.
We next show that any state in this subspace has a small excitation energy.
Then the desired statement about energy eigenstates follows immediately.
Again take arbitrary $c_1,\ldots,c_n\in\bbC$ such that $\sum_{\mu=1}^n|c_\mu|^2=1$, and let $\ket{\Gamma}=\sum_{\mu=1}^nc_\mu\ket{\Xi^{(\mu)}}$.
We first note $\sbkt{\Gamma|\Gamma}=\sum_{\mu,\zeta=1}^nc_\mu^*G_{\mu,\zeta}c_\zeta\ge1/2$.
Let $\HL'=\HL-\EGS$, which is a nonnegative operator.
We then have
\eq
\sbkt{\Gamma|\HL'|\Gamma}=\sum_{\mu=1}^n|c_\mu|^2\bkGS{(\Wl^{(\mu)})^\dagger\HL'\Wl^{(\mu)}}
+\sum_{\mu\ne\zeta}c_\mu^*c_\zeta\bkGS{(\Wl^{(\mu)})^\dagger\HL'\Wl^{(\zeta)}}.
\en
For the expectation value in the first trem we use translation invariance to see
\eq
\bkGS{(\Wl^{(\mu)})^\dagger\HL'\Wl^{(\mu)}}
=\bkGS{\Wl^\dagger\HL'\Wl}\le\frac{C}{\ell},
\en
where we used \rlb{VE}.
For the expectation value in the second term, we use the Schwarz inequality (see footnote~\ref{fn:Sch}) as
\eqa
&\bkGS{(\Wl^{(\mu)})^\dagger\HL'\Wl^{(\zeta)}}
=\bkGS{\bigl\{(\Wl^{(\mu)})^\dagger(\HL')^{1/2}\bigl\}\bigr\{(\HL')^{1/2}\,\Wl^{(\zeta)}\bigr\}}
\nl&\quad\quad\le\sqrt{
\bkGS{(\Wl^{(\mu)})^\dagger\HL'\Wl^{(\mu)}}\bkGS{(\Wl^{(\zeta)})^\dagger\HL'\Wl^{(\zeta)}}
}\le\frac{C}{\ell}.
\ena
By again using \rlb{cc}, we thus see
\eq
\sbkt{\Gamma|\HL'|\Gamma}\le\frac{C}{\ell}+(n-1)\frac{C}{\ell}=\frac{nC}{\ell},
\en
which means
\eq
\frac{\sbkt{\Gamma|\HL|\Gamma}}{\sbkt{\Gamma|\Gamma}}-\EGS\le\frac{2nC}{\ell}
\en
holds for any $\ket{\Gamma}$ in this subspace.~\qedm

\section{Extensions}
\label{s:extensions}
The Lieb-Schultz-Mattis argument applies to essentially any one-dimensional quantum system with  global $U(1)$ symmetry and translation invariance.
Here we shall briefly discuss extensions of our results (and derivations) to quantum spin chains and certain electron systems.

\subsection{Quantum spin chains}
\label{s:spin}
We treat a general quantum spin chain with arbitrary anisotropy in the z-direction.
We can also include the Dzyaloshinskii-Moriya interaction or the scalar chirality term.

Take the same one-dimensional lattice with sites $j=1,2,\ldots,pL$.
The period $p$ and the range $r$ have exactly the same meanings as before.
For each site $j$, we associate a quantum spin described by the spin operators $\hbS_j=(\hS_j^{(\mrx)},\hS_j^{(\mry)},\hS_j^{(\mrz)})$ with $(\hbS_j)^2=S_j(S_j+1)$ where the spin quantum number $S_j=1/2,1,3/2,\ldots$ satisfies the periodicity $S_j=S_{j+p}$.
We can treat any short ranged translationally invariant Hamiltonian which is invariant under any global rotation around the z-axis, or, equivalently, which commutes with the total spin operator $\hS^{(\mrz)}_{\rm tot}=\sum_{j=1}^{pL}\hS^{(\mrz)}_j$.
A general form (which is not yet the most general) is
\eqa
\HL&=\frac{1}{2}\sum_{j,k=1}^{pL}
\Bigl\{J_{j,k}(\hS_j^{(\mrx)}\hS_k^{(\mrx)}+\hS_j^{(\mry)}\hS_k^{(\mry)})
+\tJ_{j,k}(\hS_j^{(\mrx)}\hS_k^{(\mry)}-\hS_j^{(\mry)}\hS_k^{(\mrx)})\Bigr\}
+\sum_{j=1}^{pL}\hV_j
\nl
&=\frac{1}{4}\sum_{j,k=1}^{pL}
\bigl\{(J_{j,k}+i\tJ_{j,k})\,\hS_j^{+}\hS_k^{-}+(J_{j,k}-i\tJ_{j,k})\hS_j^{-}\hS_k^{+}\bigr\}
+\sum_{j=1}^{pL}\hV_j,
\lb{Hspin}
\ena
where $\hS^\pm_j=\hS_j^{(\mrx)}\pm i\hS_j^{(\mry)}$.
The term with $\tJ_{j,k}$ represents the Dzyaloshinskii-Moriya interaction.\footnote{
Note that this interaction can be written as $\boldsymbol{D}_{j,k}\cdot(\hbS_j\times\hbS_k)$ with $\boldsymbol{D}_{j,k}=(0,0,\tJ_{j,k})$.
Such an interaction which is not invariant under inversion $(j,k)\to(k,j)$ was not treated in the earlier works \cite{LSM,AL,OYA,YOA}.
The extension was made possible by the work of Koma \cite{Koma}.
See also \cite{NMT}.
It is also possible to include the scalar chirality term $J'_{j,k,\ell}\hbS_j\cdot(\hbS_k\times\hbS_\ell)$.
}
Here the exchange interaction constants satisfy the symmetry $J_{j,k}=J_{k,j}\in\bbR$, the periodicity $J_{j,k}=J_{j+p,k+p}$, and $J_{j,k}=0$ if $j=k$ or $|j-k|>r$.
The Dzyaloshinskii-Moriya interaction constants $\tJ_{j,k}$ satisfy the same constraints.
$\hV_j$ is an arbitrary hermitian operator which depends only on $\hS^{(\mrz)}_j,\ldots,\hS^{(\mrz)}_{j+r}$ and satisfies $\hV_{j+p}=\hV_j$.
A typical example, with $p=1$, is the Heisenberg antiferromagnetic chain with uniaxial anisotropy under a uniform magnetic field described by
\eq
\HL=J\sum_{j=1}^{pL}\hbS_j\cdot\hbS_{j+1}+\sum_{j=1}^{pL}\bigl\{D(\hS_j^{(\mrz)})^2-H\hS_j^{(\mrz)}\bigr\},
\lb{HD}
\en
where $(\hbS_j)^2=S(S+1)$ for all $j$ with $S=1/2,1,\ldots$.
The original work of Lieb, Schultz, and Mattis treated the model with $S=1/2$ and $D=H=0$.

Let $\GS$ be a translation invariant ground state of $\HL$ which is also an eigenstate of $\hS^{(\mrz)}_{\rm tot}$ with eigenvalue $M$.\footnote{
For a large class of antiferromagnetic chains with vanishing magnetic field, including the model \rlb{HD} with $H=0$ and even $L$, one can show that the ground state is unique and belongs to the sector with $M=0$.
See \cite{LSM,LM}, section~2.1 of \cite{AL}, and Remark~2 in section~3.1 of \cite{KeT}.
}
We then define the ``filling factor'' by $\nu=(M/L)+\sum_{j=1}^pS_j$, which satisfies $0<\nu<2\sum_{j=1}^pS_j$.

To make connection between the problems of particles and spins, we identify the number operator $\hn_j$ with $\hS^{(\mrz)}_j+S_j$.
Then all the results and derivations in sections~\ref{s:main} and \ref{s:proof} are automatically extended to the present problem.
Note that the local density operator $\rl$ now becomes
\eq
\rl=\frac{1}{\ell}\sum_{j=1}^{p\ell}(\hS^{(\mrz)}_j+S_j),
\en
which is the local magnetization (in the z direction) with the additive constant which makes $\rl\ge0$.
Thus the condition about the smallness of $\bkGS{(\Drl)^2}$ in Corollary~1 or Theorem~1 refers to the fluctuation of local magnetization (which is related to the susceptibility).
The twist operator, which is the key of the whole Lieb-Schultz-Mattis argument, becomes
\eq
\Ul=\exp\bigl[i\sum_{j=1}^{p\ell}\theta_j(\hS^{(\mrz)}_j+S_j)\bigr].
\en
This operator rotates the $j$-th spin by $\theta_j$ around the z-axis.\footnote{
In the variational estimate as in Lemma~1, it is easier to use the representation of the Hamiltonian \rlb{Hspin} in terms of the $\hS^\pm_j$ operators, and use the relation corresponding to \rlb{eceene}.
}

It is worth noting that all the trial states and the corresponding energy eigenstates lie in the Hilbert space where the eigenvalue of $\hS^{(\mrz)}_{\rm tot}$ is fixed to $M$.
Such a restriction is unnecessary from a physical point of view since there can be excitations which change the eigenvalue of $\hS^{(\mrz)}_{\rm tot}$.

\subsection{Electron systems}
\label{s:electron}
We shall discuss two types of tight-binding electron systems.
One is the class of standard lattice electron systems such as the Hubbard model, and the other is a class of hybrid models where the positions of the lattice sites are also treated as quantum mechanical degree of freedom.

We start from lattice electron systems as in the original work of Yamanaka, Oshikawa, and Affleck \cite{YOA}.
The extension is straightforward.
We again take the same lattice with $pL$ sites.
For each $j=1,\ldots,Lp$ and $\sigma=\up,\downarrow$, we let $\hc_{j,\sigma}$, $\hcd_{j,\sigma}$, and $\hn_{j,\sigma}=\hcd_{j,\sigma}\hc_{j,\sigma}$ be the annihilation, creation, and number operators, respectively, of an electron at site $j$ with spin $\sigma$.
We consider the Hamiltonian
\eq
\HL=-\sumtwo{j,k=1,\ldots,pL}{\sigma=\up,\downarrow}t_{j,k}\hcd_{j,\sigma}\hc_{k,\sigma}+\sum_{j=1}^{pL}\hV_j,
\en
where the hopping amplitude $t_{j,k}$ satisfies the same properties as before.
The interaction $\hV_j$ is an arbitrary Hermitian operator which depends only on $\hn_j,\ldots,\hn_{j+r}$, where $\hn_j=\hn_{j,\up}+\hn_{j,\dn}$, and satisfies the periodicity $\hV_j=\hV_{j+p}$.
The simplest example is the on-site Coulomb interaction $\hV_j=U\hn_{j}(\hn_{j}-1)/2=U\hn_{j,\up}\hn_{j,\dn}$, with which the model becomes the Hubbard model.

We again fix an arbitrary filling factor $\nu$ such that $0<\nu<p$, and set $N=[\nu L]$.
We then consider the Hilbert space with $N$ electrons with spin $\up$ and $N$ electrons with spin $\dn$.  Note that there are $2N$ electrons.
Then all the results and derivations in sections~\ref{s:main} and \ref{s:proof} extend to the present problem if we replace $\hn_j$ in the spinless model with $\hn_{j,\up}$ (or $\hn_{j,\dn}$).
Thus the second condition in Corollary~1 now deals with the fluctuation of the density of up-spin electrons.
The twist operator, for example, becomes
\eq
\Ul:=\exp\bigl[i\sum_{j=1}^{p\ell}\theta_j\,\hn_{j,\up}\bigr],
\lb{Ulup}
\en
where only up-spin electrons are modified.

It is obvious that all the results can be extended to various lattice models in which electrons couple to fixed spin degrees of freedom.
See \cite{YOA} for discussion about the Kondo-Heisenberg model.

\bigskip
Next, to demonstrate the generality of the Lieb-Schultz-Mattis method, we briefly discuss a model in which the positions of the lattice sites (i.e., the locations of ions forming the lattice) are treated as quantum mechanical degrees of freedom.
The model is capable of describing, e.g., the Peierls instability.

Consider a one-dimensional lattice whose sites are labelled as $j=1,\ldots,pL$.
Let $\hbr_j$ denote the displacement of the lattice site $j$ from its equilibrium position, and $\hbp_j$ be the corresponding momentum.
Then our Hamiltonian is
\eq
\HL=\sum_{j=1}^{pL}\frac{\hbp_j^2}{2M_j}+V'(\hbr_1,\ldots,\hbr_{pL})
-\sumtwo{j,k=1,\ldots,pL}{\sigma=\up,\downarrow}t_{j,k}(\hbr_j,\hbr_k)\,\hcd_{j,\sigma}\hc_{k,\sigma}+\sum_{j=1}^{pL}\hV_j,
\en
where the first two terms describe the dynamical degrees of freedom of lattice sites, and the remaining two terms describe that of tight-binding electrons.
Here $M_j$ denotes the mass of the $j$-th ion, and $V'(\hbr_1,\ldots,\hbr_{pL})$ is the interaction potential.
The electron part is  almost the same as the previous model except that the hopping amplitude $t_{j,k}(\hbr_j,\hbr_k)=\{t_{k,j}(\hbr_k,\hbr_j)\}^\dagger$ can now depend on the position operators $\hbr_j$ and $\hbr_k$, and the interaction $\hV_j$ is an arbitrary Hermitian operator which depends only on $\hn_j,\ldots,\hn_{j+r}$ and $\hbr_j,\ldots,\hbr_{j+r}$.
We assume that everything is defined so that the Hamiltonian becomes invariant under the translation by $p$ sites.

The filling factor $\nu$ and the Hilbert space for electrons are defined in exactly the same manner as the lattice model treated above.
We then define $\GS$ to be a translation invariant ground state of the whole system including the lattice positions and the electrons.
Then, rather surprisingly, all the results in section~\ref{s:main} and their proofs in section~\ref{s:proof} can be extended to the present (rather complicated) model with little modification.

Let us see some of the crucial steps in the proof.
The twist operator is of course defined as \rlb{Ulup}, where only the degrees of freedom of up-spin electrons are modified.
Then the energy estimate \rlb{DE2}  in the proof of Lemma~1 is modified as
\eqa
\DE&\le\sum_{j,k=1}^{pL}2\bigl\{1-\cos(\theta_j-\theta_k)\bigr\}\,\sbkt{t_{j,k}(\hbr_k,\hbr_j)\,\hcd_{j,\up}\hc_{k,\dn}}
\nl&\le
\sum_{j,k=1}^{pL}2\bigl\{1-\cos(\theta_j-\theta_k)\bigr\}\,\sqrt{\bbkt{\{t_{j,k}(\hbr_k,\hbr_j)\}^\dagger\, t_{j,k}(\hbr_k,\hbr_j)}},
\lb{DE3B}
\ena
where we used the Schwarz inequality.
It is reasonable to assume that the expectation value $\bkt{\{t_{j,k}(\hbr_k,\hbr_j)\}^\dagger\, t_{j,k}(\hbr_k,\hbr_j)}$ is bounded by some constant.
The rest of the proof is essentially the same as before.
Note in particular that the proof of Lemma~2 about near orthogonality is carried out as it is because the argument involves only the $U(1)$ phase of up-spin electrons.

\bigskip
{\small
I wish to thank Tohru Koma, Masaki Oshikawa, and Haruki Watanabe for valuable discussions which were essential for the present work, and Ian Affleck, Hosho Katsura, Tomonari Mizoguchi, Lee SungBin, Akinori Tanaka, Masafumi Udagawa, and Masanori Yamanaka for useful discussions and correspondences.
The present work was supported by JSPS Grants-in-Aid for Scientific Research no.~16H02211.
}


\end{document}